\documentclass[pra,aps,showpacs,floatfix,twocolumn]{revtex4-1}
\usepackage{graphicx}

\begin{document}

\title{Mass-imbalance induced metal-insulator transition \\ in a three-component Hubbard model}

\author{Duong-Bo Nguyen, Duy-Khuong Phung, Van-Nham Phan, and Minh-Tien Tran}
\affiliation{Institute of Research and Development, Duy Tan University, K7/25 Quang Trung, Danang, Vietnam }
\affiliation{ Institute of Physics, Vietnam Academy of Science and Technology, 10 Dao Tan,
Hanoi, Vietnam}

\begin{abstract}
The effects of mass imbalance in a three-component Hubbard model are studied
by the dynamical mean-field theory combined with exact diagonalization. The model describes a fermion-fermion mixture of two different particle species with a mass imbalance. One species is two-component fermion particles, and the other is single-component ones. The local interaction between particle species is considered isotropically. It is found that the mass imbalance can drive the mixture from insulator to metal. The insulator-metal transition is a species-selective-like transition of lighter mass particles and occurs only at commensurate particle densities and moderate local interactions. For weak and strong local interactions the mass imbalance does not change the ground state of the mixture.
\end{abstract}


\maketitle

\section{Introduction}
One of the fascinating problems in condensed matter physics is the metal-insulator transition (MIT).
Especially, the MIT driven by electron correlations has attracted a lot of attention.
The electron correlations suspend the double occupancy of electrons,
and the suspension causes the electron localization~\cite{Mott}.
With the achievements of ultracold atom techniques, optical lattices
can experimentally simulate various theoretical models of electron correlations~\cite{Bloch}. They
have been providing a novel stage for studying correlation effects in materials.
In particular, the Mott insulator was observed in optical lattices of fermionic $^{40}$K  atoms
with two hyperfine states and repulsive interaction between
them~\cite{Jordens,Schneider}. These optical lattices really simulate the two-component Hubbard model and they provide a connection between experimental observations and theoretical predictions.
The optical lattices can also be established with different particle species which can
be extended to have both large hyperfine multiplets and mass imbalances. For example,
a mixture of single-spin state $^{40}$K immersed in
two-component fermionic atoms $^6$Li or a mixture of two-component
state $^{171}$Yb and six-component state $^{173}$Yb have
already been achieved~\cite{Spiegelhalder,Taie}.
Such achievements provide
possible realizations of the MIT in multi-component
correlation systems. Theoretical studies already predicted a MIT in
three-component Hubbard models~\cite{Gorelik, Inaba}.
The MIT occurs at commensurate particle densities when the local
interaction is isotropic~\cite{Gorelik}.
With an anisotropy of
the local interactions, the MIT
is also found at incommensurate half filling~\cite{Inaba}.
However, in these studies, all component particles have the same
mass. Experiments can also separately tune the individual effective mass of each particle species and establish imbalanced mass mixtures. Indeed, in a mixture of $^6$Li and $^{40}$K atoms,
the mass imbalance can be tuned in a wide range~\cite{Dalmonte}.
The mixtures of two-component particles with different masses in the optical lattice generally lead to the difference between the hopping amplitudes associated with each component. With deviation from the balance limitation, some phase transitions might happen and change the ground state. Low-temperature properties of the optical lattice in the influence of the imbalance therefore are important and need to be considered. Indeed, mass imbalance in the two-component Hubbard model has been studied in detail in the one-dimensional and three dimensional cases that clarified the competing of some order states such as superfluid, charge-, and
spin-ordered states in the optical lattices~\cite{Dao07,Caza,Take,Dao}.
Increasing the hopping imbalance or increasing difference between the bandwidth of each particle species might lead the systems to a situation that one species is transformed from the metallic to the insulating state due to  electron-electron correlations while the other still remains in its metallic or insulating state. The transition is similar to the orbital-selective MIT in correlated multiorbital systems~\cite{Anisimov,Koga,Liebsch2004,Medici}. However, in ultracold multicomponent mixtures the MIT may deal with odd numbers of components, whereas in multiorbital systems it is impossible~\cite{Gorelik,Inaba}. Therefore,
studying the effects of massimbalance on the MIT therefore is an interesting subject of the optical lattice systems.
Actually, the effects of the mass imbalance
on the MIT in fermion mixtures of two single-component particle species have been studied~\cite{Dao07,Caza,Take,Dao}.
The two species show distinct properties, which deviate from the behaviors of balanced mass mixtures~\cite{Dao}.
In fermion-fermion mixtures of multicomponent particle species with extreme mass imbalance, where
one particle species is extremely heavy and immobile,
distinct MITs were also found~\cite{BoTien}. The MIT can occur at both commensurate and incommensurate particle
densities. Despite that one species is always localized, the MIT can be classified as a collective, species-selective or inverse transition, depending on the local interaction
and particle densities~\cite{BoTien}.

In this paper we study the effects of the mass imbalance on the MIT in
multicomponent fermion-fermion mixtures. The mixtures consist of two-component
and single-component fermion particles.
In contrast to the previous study~\cite{BoTien},
both particle species have finite masses and mobilities.
Such fermion-fermion mixtures can be realized by
$^{40}$K with $^6$Li atoms, or light $^6$Li or $^{40}$K with
heavy fermionic isotopes of Sr or Yb.
To model these mixtures with a mass imbalance we propose a three-component
Hubbard model with different hopping amplitudes.
Actually, the hopping amplitudes can experimentally be tuned by
the lattice potential and the recoil energy~\cite{Zwerger}.
In the balanced mass mixtures, the local interaction can drive
the mixtures from metal to insulator states at commensurate particle densities~\cite{Gorelik}.
In the extreme case, where one species is completely localized or nonhopping, the Hubbard model above reduces to a three-component Falicov-Kimball model and
the MIT is found existing in both states of commensurate and incommensurate particle densities~\cite{BoTien}. Studying the MIT in between, i.e., with the hopping
imbalance in the systems, therefore seems to be important to release the low-temperature quantum properties under the competition between the complex of electronic
kinetic energy and the strong correlations.

We study the three-component Hubbard model by employing dynamical mean-field theory (DMFT)
with exact diagonalization (ED).
The DMFT has been used successfully to study the strongly correlated electron
systems~\cite{Metzner,GKKR}. The previous
studies of MIT in the three-component Hubbard and Falicov-Kimball models were also
based on the DMFT~\cite{Gorelik,Inaba,BoTien}.
The DMFT is exact in infinite dimensions and fully captures
local dynamical fluctuations~\cite{Metzner,GKKR}. However, it loses nonlocal
correlations in finite dimensions. Within the DMFT,
we found a MIT that is solely
driven by the mass imbalance. This transition is
a species-selective-like transition of lighter particles and occurs only at commensurate total
particle densities and moderate local interactions. For weak and strong correlations,
the mass imbalance cannot drive the mixture out of its state.

The present paper is organized as follows. In Sec. II we describe
the three-component Hubbard model with a mass imbalance.
In this section we also present the application of the DMFT to the proposed Hubbard model.
Numerical results for the detected MIT is presented in Sec. III.
Finally,
the conclusion is presented in Sec. IV.

\section{Three-component Hubbard model and its dynamical mean field theory}

We consider a three-component Hubbard model, the Hamiltonian of which reads
\begin{eqnarray}
H &=& - \sum_{<i,j>,\alpha} t_\alpha c^{\dagger}_{i\alpha} c_{j\alpha} + \frac{U}{2} \sum_{i,\alpha \neq \alpha'}
n_{i\alpha} n_{i\alpha'},  \label{ham}
\end{eqnarray}
where $c^{\dagger}_{i\alpha}$ ($c_{i\alpha}$) is the creation
(annihilation) operator for the fermionic particle with hyperfine
multiplet $\alpha$ at site $i$. $\alpha$ takes three different
values, for instance, $\alpha=1, 2, 3$. $n_{i\alpha} = c^{\dagger}_{i\alpha} c_{i\alpha}$
is the number operator of the $\alpha$-component
fermionic particles at site $i$.
$t_\alpha$ is the hopping parameter of the $\alpha$-component
fermionic particles. $U$ is the local
interaction between the three component states of particles.
A common
chemical potential $\mu$ is also introduced to control the total
particle density $n=\sum_{i\alpha} \langle n_{i\alpha} \rangle /N$, where $N$ is the number of lattice sites.
The three-component Hubbard model can be realized by loading ultracold fermionic atoms with three hyperfine multiplets
or fermion-fermion mixtures of different atomic species into optical lattices. However, in the Hamiltonian in Eq. (\ref{ham}),
the trapping potentials in the optical lattices are not considered.

The mass imbalance solely depends on the hopping parameters $t_\alpha$. In the three-component
Hubbard model, the mass imbalance actually means the difference of the hopping parameters. In optical
lattices, the particle hopping is established by the particle tunneling
between nearest neighbor lattice potential wells. It can be tuned by the lattice potential amplitude $V^{\rm lat}_\alpha$
and the recoil energy $E_{r\alpha}$ of each component state of particles \cite{Zwerger}
\begin{eqnarray}
t_\alpha \approx \frac{4}{\sqrt{\pi}} E_{r\alpha} v_\alpha^{3/4} \exp(-2\sqrt{v_\alpha}) ,
\end{eqnarray}
where $v_\alpha = V_\alpha^{\rm lat} /E_{r\alpha}$. The recoil energy $E_{r\alpha}=k^2/2 m_\alpha$,
where $k$ is the wave number of the laser forming the optical lattice, and $m_\alpha$ is the mass
of the $\alpha$- component particles. The lattice potential amplitude $V^{\rm lat}_\alpha$ can be different
for different hyperfine states of particles. As a result, the hopping parameters can also be different even
for the hyperfine states of the same particles with identical masses. In the following, we will consider
the hopping imbalance $t_1 \neq t_2 = t_3$. This case can be interpreted as a fermion-fermion mixture of two different
particle species. One species is particles with a single hyperfine state ($\alpha=1$), while the other is particles with two hyperfine states ($\alpha=2,3$).
Such mixture can be realized by loading fermion
atoms $^{40}$K and $^6$Li, or of light atoms $^6$Li or $^{40}$K with heavy fermion isotopes of Sr or Yb,
into optical lattices.
We parametrize the hopping amplitudes by
\begin{eqnarray}
t &=& \frac{t_1+t_2}{2} , \\
\Delta t &=& \frac{t_2 - t_1}{t_1+t_2}\label{Dt} .
\end{eqnarray}
$t$ is the average hopping amplitude of two particle species, and $\Delta t$ describes the mass
imbalance between them.
It is clear that $-1 \leq \Delta t \leq 1$. $\Delta t=\pm 1$ are the extreme mass imbalance,
where one particle species is extremely heavy and localized~\cite{BoTien}. Actually, the effective mass of the particles is inversely proportional to their hopping parameter.
$\Delta t >0$ indicates that the two-component particles are lighter than the single-component particles,
and vice versa for $\Delta t <0$.
$\Delta t=0$ is the balanced mass mixture and the model corresponds
to the isotropic three-component Hubbard model~\cite{Gorelik}.
In the balanced mass case,
with sufficiently strong local
interaction,  the Mott insulator exists at commensurate particle
densities~\cite{Gorelik}.
The mass imbalance parameter $\Delta t$ can experimentally be tuned in a wide range.
For instance, in a mixture of $^6$Li and $^{40}$K atoms, $\Delta t$ can vary from $0.3$
to $0.85$~\cite{Dalmonte}.

We study the three-component Hubbard model by employing the DMFT. Within the DMFT the self-energy
is a local function of frequency. It is exact in infinite dimensions.
However, in finite dimensions, the DMFT neglects nonlocal correlations.
The DMFT is well described in the literature, for example, in Ref.~\onlinecite{GKKR}. For self-contained purposes, we present here a description of applying the DMFT to the three-component
Hubbard model.
The Green's function of the $\alpha$-component particles
reads
\begin{equation}
G_\alpha(\mathbf{k},i\omega_n) = \displaystyle\frac{1}{i\omega_n+\mu + t_\alpha \varepsilon_{\mathbf{k}}-\Sigma_\alpha(i\omega_n)},
\end{equation}
where $\omega_n$ is the Matsubara frequency,
$\varepsilon_{\mathbf{k}} = \sum_{<i,j>} \exp(i \mathbf{k} \cdot (\mathbf{r}_i - \mathbf{r}_j)) $ is the lattice structure factor, and
$\Sigma_\alpha(i\omega_n)$ is the self-energy.
The local self-energy is determined from the dynamics of a
single three-component particle embedded in a dynamical mean field. The action of the effective single particle reads
\begin{eqnarray}
\mathcal{S}_{{\rm imp}} &=&  - \int\limits_{0}^{\beta}
\int\limits_{0}^{\beta} d\tau d\tau' \sum_{\alpha}
c^{\dagger}_{\alpha}(\tau) \mathcal{G}^{-1}_\alpha(\tau-\tau')
c_{\alpha}(\tau') \nonumber \\
&& + \frac{U}{2} \int\limits_{0}^{\beta} d\tau \sum_{\alpha \neq \alpha'} n_{\alpha}(\tau) n_{\alpha'}(\tau) ,
\label{action}
\end{eqnarray}
where $\mathcal{G}_\alpha(\tau)$ is a Green's function which represents the dynamical
mean field. $\mathcal{G}_\alpha(\tau)$ plays as the bare Green's function in relation to the local Green function.
It relates to the self-energy and the local Green's function by the
Dyson equation
\begin{equation}
\mathcal{G}^{-1}_\alpha(i\omega_n) = G^{-1}_\alpha(i\omega_n) + \Sigma_\alpha(i\omega_n) .
\label{dyson}
\end{equation}
Here, the local Green's function is
\begin{equation}
G_\alpha(i\omega_n) = \int d\varepsilon \rho_{0}(\varepsilon)
\displaystyle\frac{1}{i\omega_n+\mu-\Sigma_\alpha(i\omega_n)+t_\alpha \varepsilon},
\label{local}
\end{equation}
where $\rho_{0}(\varepsilon)=\sum_{\mathbf{k}}
\delta(\varepsilon-\varepsilon_{\mathbf{k}})$ is the bare density
of states (DOS). Without loss of generality, we use the
semicircular DOS
\begin{equation}
\rho_0(\varepsilon) = \frac{2}{\pi} \sqrt{1-\varepsilon^2} .
\end{equation}
With the semicircular DOS, from Eqs. (\ref{dyson}) and (\ref{local})  we immediately obtain
the Green's function representing the dynamical mean field $\mathcal{G}(i\omega_n)$ from
the local Green's function \cite{GKKR}
\begin{equation}
\mathcal{G}^{-1}_\alpha(i\omega_n) = i\omega_n + \mu - \frac{t_\alpha^2}{4} G_\alpha(i\omega_n) .
\end{equation}
The self-consistent condition of the DMFT requires that the Green's function
obtained from the effective action in Eq. (\ref{action}) be identical to the local Green's function in Eq. (\ref{local}); i.e.,
\begin{equation}
G^{{\rm imp}}_\alpha(i\omega_n) = G_\alpha(i\omega_n) .
\end{equation}
This equation completes the set of self-consistent equations for
the Green's function. It can be solved numerically by
iterations~\cite{GKKR}. The most time-consuming part is the
solving of the action $\mathcal{S}_{\rm imp}$ in Eq. (\ref{action}).
There are several ways to calculate the Green's function from the action $\mathcal{S}_{\rm imp}$~\cite{GKKR}.
Here, we employ an ED method to calculate it~\cite{GKKR,Krauth}. The action in Eq.
(\ref{action})  is
essentially equivalent to the Anderson impurity model~\cite{GKKR,Krauth}
\begin{eqnarray}
\lefteqn{
H_{\rm AIM} = -\mu \sum_{\alpha} c^{\dagger}_{\alpha} c_{\alpha} +
\frac{U}{2} \sum_{\alpha \neq \alpha'} n_{\alpha} n_{\alpha'} } \nonumber \\
&&+ \sum_{p,\alpha} V_{p\alpha} a^{\dagger}_{p\alpha} c_{\alpha} + {\rm H.c.}
+ \sum_{p,\alpha} E_{p\alpha} a^{\dagger}_{p\alpha} a_{p\alpha} ,
\label{anderson}
\end{eqnarray}
where $a^{\dagger}_{p\alpha}$ ($a_{p\alpha}$) is the creation
(annihilation) operator which represents a conduction bath with
energy level $E_{p\alpha}$. $V_{p\alpha}$ is the coupling of the conduction
bath with the impurity. The connection between  the Anderson
impurity model in Eq. (\ref{anderson}) and the action in Eq.
(\ref{action}) is the following identity relation of the bath
parameters \cite{GKKR,Krauth}
\begin{equation}
\sum_{p} \frac{|V_{p\alpha}|^2}{i\omega_n - E_{p\alpha}} = \lambda_\alpha(i\omega_n) ,
\end{equation}
where $\lambda_\alpha(i\omega_n) = i\omega_n + \mu - \mathcal{G}^{-1}_\alpha(i\omega_n)$.
The ED limits the conduction bath to finite $n_s-1$ orbits ($p=1,2,...,n_s-1$).
Then  $\lambda(i\omega_n)$ is approximated by
\begin{equation}
\lambda^{(n_s)}_{\alpha}(i\omega_n) = \sum_{p=1}^{n_s-1} \frac{|V_{p\alpha}|^2}{i\omega_n - E_{p\alpha}} .
\end{equation}
The bath parameters are determined from minimization of the distance $d$
between $\lambda_\alpha(i\omega_n)$ and $\lambda^{(n_s)}_{\alpha}(i\omega_n)$,
\begin{equation}
d = \frac{1}{M+1} \sum_{n=0}^{M}
\omega_n^{-k} | \lambda_{\alpha}(i\omega_n)-\lambda^{(n_s)}_{\alpha}(i\omega_n) |^2 ,
\end{equation}
where $M$ is a large upper cutoff of the Matsubara
frequencies~\cite{GKKR,Krauth}. The parameter $k$ is introduced to
improve the minimization at low Matsubara frequencies.
In particular, we take $k=1$ in the
numerical calculations.  When the bath
parameters are determined, we calculate the Green's function of the
Anderson impurity model in Eq. (\ref{anderson}) by
ED~\cite{GKKR,Krauth}. We also calculate the inter-species double occupancy  $D_{\rm inter} = \sum_{i}\langle n_{i1} (n_{i2}+n_{i3}) \rangle /N$, and the intra-species double occupancy $D_{\rm intra} = \sum_{i} \langle n_{i2} n_{i3} \rangle /N$. These double occupancies show the rate of the lattice sites occupied by two particles,
and are experimentally accessible~\cite{Jordens}.

\section{Metal-insulator transition}

\begin{figure}[t]
\includegraphics[width=0.4\textwidth]{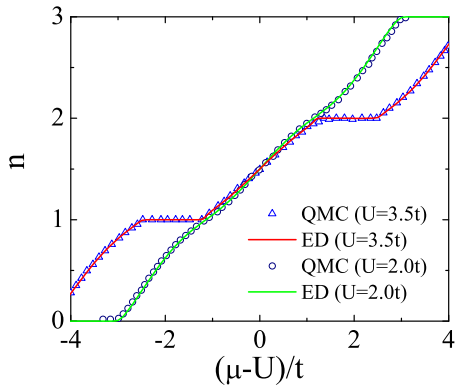}
\caption{(Color online) The total particle density $n$, calculated by DMFT+ED (lines) and by DMFT+QMC (symbols), as a function of
the chemical potential $\mu$ for different values of interaction $U$ in the  balanced mass case $\Delta t=0$
($T=0.025 t$). The DMFT+QMC results are reproduced from Ref.~\onlinecite{Gorelik}.  }
\label{figad1}
\end{figure}

\begin{figure}[b]
\includegraphics[width=0.4\textwidth]{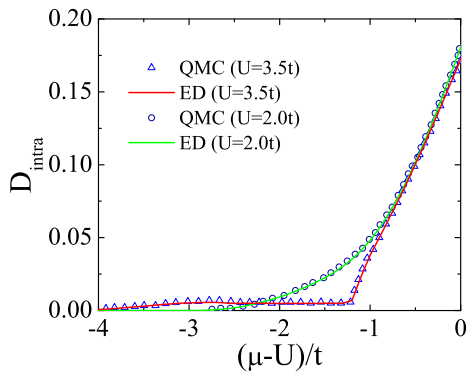}
\caption{(Color online) The double occupancy as a function of
the chemical potential $\mu$, calculated by DMFT+ED (lines) and by DMFT+QMC (symbols) for different values of interaction $U$ in the balanced mass case $\Delta t=0$
($T=0.025 t$). The DMFT+QMC results are reproduced from Ref.~\onlinecite{Gorelik}.  }
\label{figad2}
\end{figure}

\begin{figure}[b]
\includegraphics[width=0.47\textwidth]{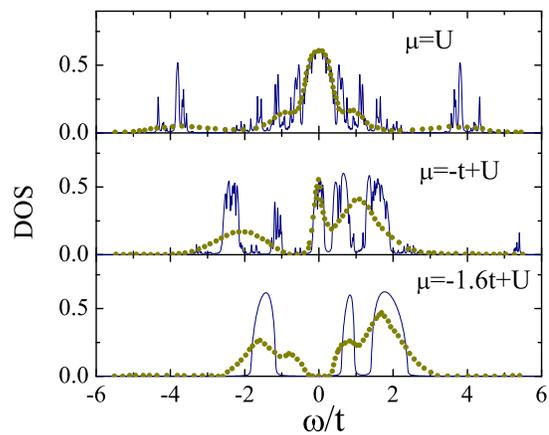}
\caption{(Color online) The density of states (DOS), calculated by DMFT+ED (lines)
and by DMFT+QMC (symbols) for different chemical potentials $\mu$ in the balanced mass case $\Delta t=0$
($U=3 t$, $T=0.05 t$). The DMFT+QMC results are reproduced from Ref.~\onlinecite{Gorelik}.  }
\label{figad3}
\end{figure}

In this section we present numerical results analyzing the MIT under the influence of the mass-imbalance indicated by $\Delta t$ [c.f. Eq.~(\ref{Dt})].
The numerical results are obtained by the DMFT+ED with $3$ bath orbits per one impurity component (i.e., $n_s=4$).
Actually, the computational time grows quickly with $n_s$, since the impurity has $3$ components.
For the single-band Hubbard model, the DMFT+ED shows that two bath levels per one impurity component usually give adequate results~\cite{Liebsch}. To check the accuracy of the ED, we compare our results with the ones obtained by the DMFT plus the Hirsch-Fye quantum Monte Carlo (QMC) simulations in the case of mass balance ($\Delta t=0$)~\cite{Gorelik}. The QMC simulations are an exact impurity solver, and unlike the ED they do not suffer the finite-size effects~\cite{GKKR}. First, Fig.~\ref{figad1} illustrates the comparison of the total particle density $n$ as a function of the chemical potential $\mu$ obtained by DMFT combining with ED and QMC for two different $U$ values. When the line $n(\mu)$ exhibits a plateau, it indicates an insulating state. The width of the plateau is equal to the insulating gap. Figure~\ref{figad1} apparently shows an excellent agreement between the ED and QMC simulations for both metallic and insulating phases in a whole range of the total particle density. In Fig.~\ref{figad2} we continue specifying the efficiency of the DMFT+ED calculation to consider the MIT in the Hubbard model by putting its beside DMFT+QMC results of the intra-species double occupancy $D_{\rm intra}$. As a function of the chemical potential $\mu$, $D_{\rm intra}$ again recovers the excellent agreement between the ED and QMC simulations for both metallic and insulating phases. The double occupancy is suppressed in the insulating phase. However, it still remains finite. The finite value of the double occupancy in the insulating phase therefore is not a finite-size effect of the ED impurity solver. As pointed out in the literature the Brinkman-Rice approximation shows that the double occupancy vanishes in the Mott insulator~\cite{Brinkman}. In reality a finite local interaction always allows virtual hoppings, that produce very small but nonzero double occupancy in the insulating state. That feature has been addressed by the DMFT~\cite{GKKR,Zhang,Rozenberg,GeorgesKrauth}. The double occupancy vanishes only at the strong-interaction limit. Actually, the Brinkman-Rice approximation is based on the Gutzwiller variational wave function and the Gutzwiller approximation for evaluating the variational ground-state energy, and it admits the vanishing of the double occupancy in the insulating phase~\cite{Brinkman,Gutzwiller}. However, at finite dimensions, without the additional Gutzwiller approximation, the Gutzwiller variational wave function always produces a finite double occupancy for any local interaction~\cite{Gebhard}. Moreover, extending DMFT for the finite-dimension case where the nonlocal correlations are taken into account has also illustrated the incomplete suppression of the double occupancy in the Mott insulator~\cite{Jarrell,Parcollet}. In experiment, suppression of the double occupancy in the MIT of the ultracold  two-component fermion atoms has been observed, but a small number of lattice sites, typically a few percent, still remains doubly occupied in the Mott insulator~\cite{Jordens}. To complete the comparison, Fig.~\ref{figad3} illustrates the density of states (DOS) of the particle component
${\rm Im} G_\alpha(\omega-i\eta)/\pi$, evaluated by both DMFT+QMC~\cite{Gorelik} and DMFT+ED. In the DMFT+ED calculation, we used $\eta=0.01t$. Despite the spiky structure in the DOS, the main features of the DOS obtained by the DMFT+ED resemble the ones shown by the DMFT+QMC. Inspecting the DOS at the chemical potential (i.e.,~the DOS at $\omega=0$) we see that the quantities determined by both ED and QMC simulations fit well together. In the metallic phase the DOS at the chemical potential level is always finite, while it vanishes in the insulating phase. From the above comparisons of ED and QMC results, we conclude that the ED solving for the effective impurity problem with $n_s=4$ in the DMFT already gives adequate results. In the following, we extend the calculation to the mass imbalance case (i.e.~$\Delta t\neq 0$) and then discuss its effects on the MIT in the three-component Hubbard model.

\begin{figure}[t]
\includegraphics[width=0.4\textwidth]{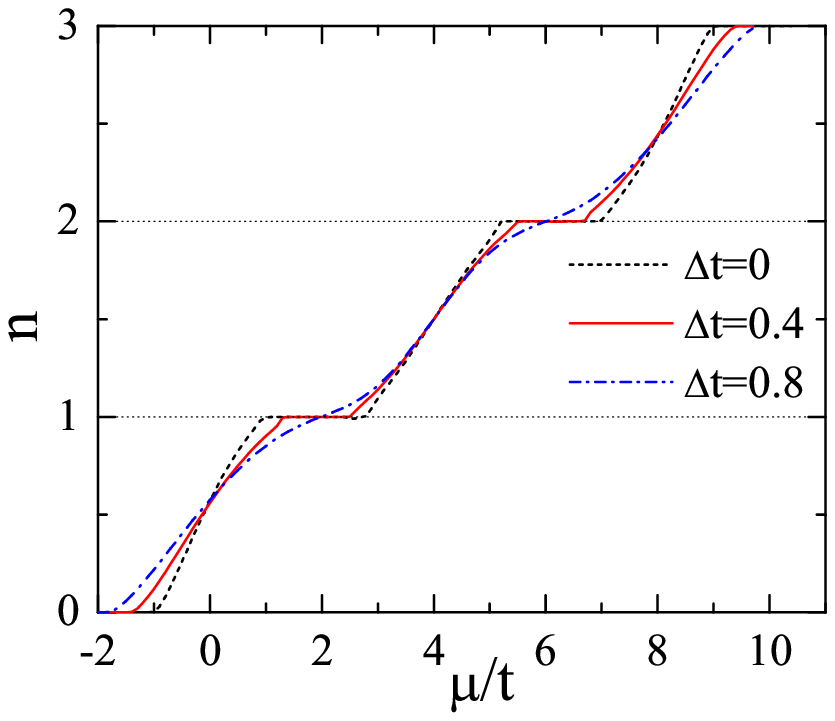}
\caption{(Color online) The total particle density $n$ as a function of
the chemical potential $\mu$ for positive mass imbalances ($\Delta t>0$)
at fixed interaction $U=4 t$ ($T=0.02t$).  } \label{fig1}
\end{figure}

In Fig.~\ref{fig1} we plot
the total particle density as a function of the chemical potential when the mass imbalance varies in the region $\Delta t > 0$ at a given isotropic local interaction $U$. The value of the local interaction is chosen that allows a MIT. In the case of $U=4t$ we observe that the mass imbalance possibly drives the mixture only from insulator to metal. Figure~\ref{fig1} also shows that when the mass imbalance is absent, $\Delta t=0$, the line $n(\mu)$ exhibits plateaus at $n=1$ and $n=2$. The mixture is in the insulating phase at the commensurate densities only~\cite{Gorelik}.
With increasing the mass imbalance $\Delta t$, the plateaus reduce and then disappear at large
mass imbalances. These behaviors show a transition from insulator to metal solely driven by the mass imbalance at the commensurate densities. Actually, when $\Delta t$ increases, the two-component particles become lighter and the single-component particles become heavier. Due to this property, the mobility of the two-component particles tends to increase too. At large mass imbalances it overcomes the local interaction, and the mixture becomes metallic. In the $\Delta t=1$ limit, the single-component particles are localized due to the vanishing of their hopping, but the two-component particles can be in the metallic phase~\cite{BoTien}.
In such way, the mass imbalance can only drive the mixture from insulator to metal. It cannot
drive the mixture out of the metallic state.

\begin{figure}[t]
\includegraphics[width=0.4\textwidth]{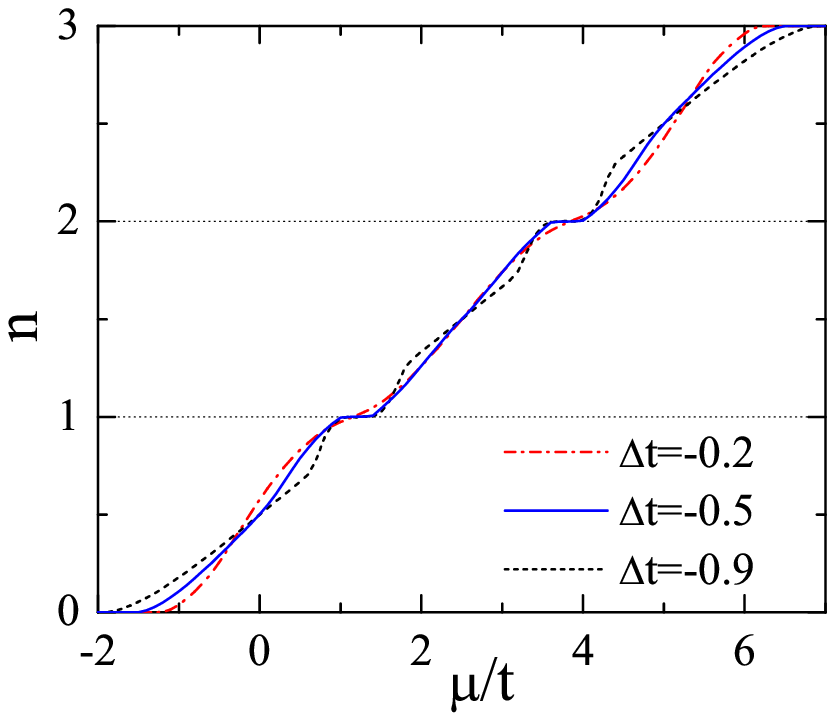}
\caption{(Color online) The total particle density $n$ as a function of
the chemical potential $\mu$ for negative mass imbalances ($\Delta t <0$)
at fixed interaction $U=2.5 t$ ($T=0.02t$).  } \label{fig1a}
\end{figure}

\begin{figure}[t]
\includegraphics[width=0.4\textwidth]{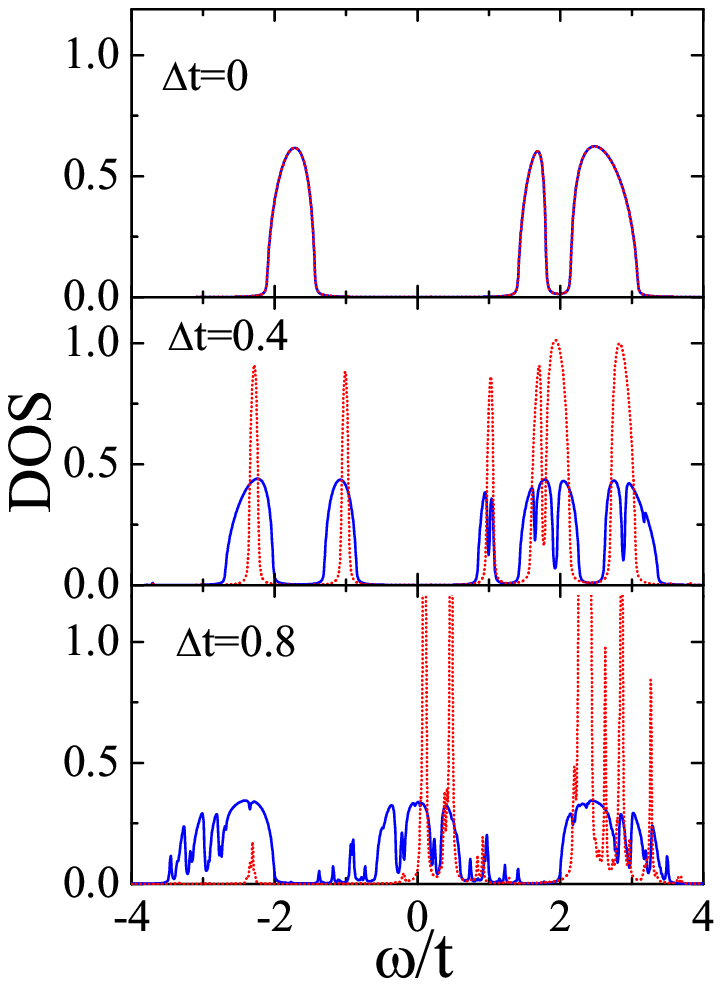}
\caption{(Color online) The density of states (DOS) of the two-component particles (blue solid lines)
and of single-component particles (red dotted lines) for positive mass imbalances $(\Delta t >0)$
at fixed total particle density $n=1$ and interaction $U=4 t$ ($T=0.02t$). }
\label{fig2}
\end{figure}

Next, we continue considering a dependence of the total particle density on chemical potential but in the opposite region of the mass imbalance ($\Delta t<0$). Figure~\ref{fig1a} shows us that for a given local interaction, $U=2.5t$, the mass imbalance can only drive the mixture from metal to insulator.
With the increasing of $|\Delta t|$, the two-component particles  become heavier, and the single-component particles become lighter. When $|\Delta t|$ remains being small, $n(\mu)$ does not show any plateau and the system is in the metallic state. The plateaus indicating an insulating state only appear if $|\Delta t|$ is larger than a critical value. That happens at the commensurate densities $n=1$ and $n=2$, similar to the opposite situation with $\Delta t>0$ (see Fig.~\ref{fig1}). In the limit of $\Delta t=-1$, the two-component particles become localized due to the vanishing of their hopping. However, the hopping of single-component particles remains finite. For weak interspecies interactions, the single-component particles are in a metallic phase. However, with increasing the interspecies interaction, the band of the single-component particles is split into two subbands, separated by a gap. This constitutes a MIT driven by inter-species particle correlations.
Actually, the model with $\Delta t=-1$ is equivalent to the spinless Falicov-Kimball model for strong interactions~\cite{FreericksZlatic}. The spinless Falicov-Kimball model also exhibits a Mott-like MIT by splitting the conduction band into two subbands~\cite{FreericksZlatic,Dongen92,Dongen}.

\begin{figure}[t]
\includegraphics[width=0.4\textwidth]{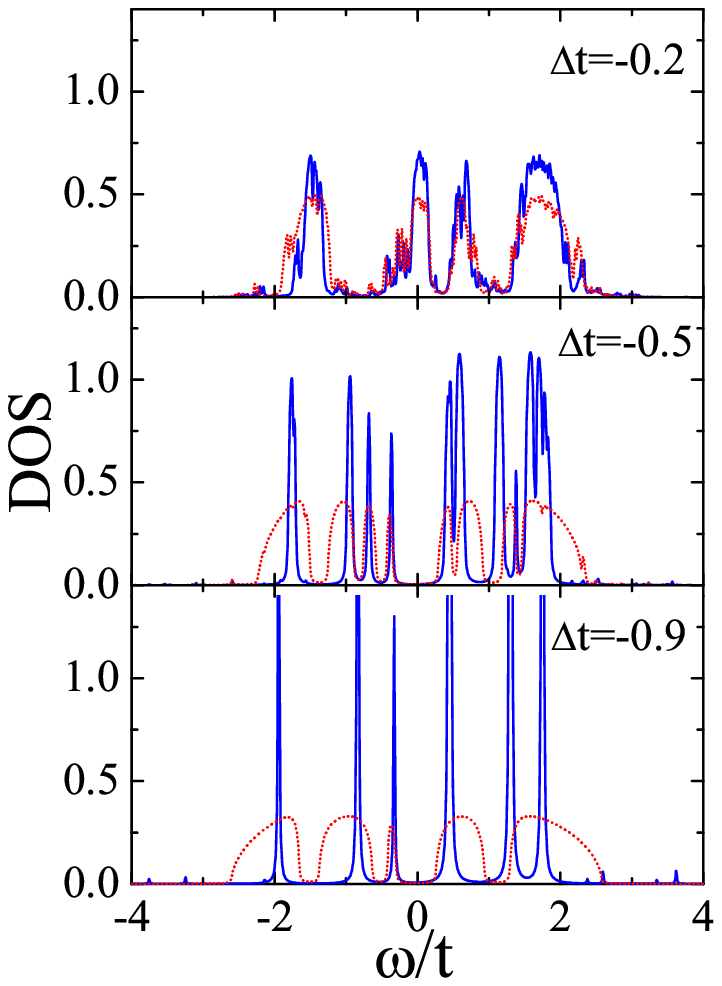}
\caption{(Color online) The density of states (DOS) of the two-component particles (blue solid lines)
and of single-component particles (red dotted lines) for negative mass imbalances $(\Delta t <0)$
at fixed total particle density $n=1$ and interaction $U=2.5 t$ ($T=0.02t$). }
\label{fig2b}
\end{figure}

The MIT can be understood in more detail in a feature of the DOS. In Figs.~\ref{fig2} and \ref{fig2b} we plot the DOS of particles
at the commensurate density $n=1$. The other commensurate density $n=2$ is equivalent to $n=1$ via the particle-hole symmetry.
Figures~\ref{fig2} and \ref{fig2b} confirm again the MIT driven by the mass imbalance in both regions $\Delta t>0$ and $\Delta t<0$.
When the mass imbalance is absent, $\Delta t=0$,
particles of two
species equally participate in the MIT driven by the local interaction~\cite{Gorelik}.
However, as addressed before in Fig.~\ref{fig1} and Fig.~\ref{fig1a}, in the region of $\Delta t >0$, the mass imbalance can drive the mixture only from insulator to metal, whereas in the region of $\Delta t <0$, it only drives the mixture from metal to insulator.
In the $\Delta t>0$ region, with increasing $\Delta t$ the two-component particles become lighter, and
the bands are mostly occupied by them. This indicates that the mass imbalance also induces a population imbalance between
the two particle species.
In the extreme limit, $\Delta t=1$, the single-component particles are localized, and
only the two-component particles take part in the MIT~\cite{BoTien}.
We can refer to the MIT driven by the mass imbalance as a species-selective-like MIT. In this MIT the lighter particles are dominantly
involved in driving the mixtures to the metallic state. However, this species-selective-like MIT is different in comparison with the orbital-selective MIT~\cite{Anisimov,Koga,Liebsch2004,Medici}. In the orbital-selective MIT, the wide band remains metallic when the narrow band becomes insulating.
In this selective-species-like MIT, the bands of both species are insulating and when the MIT occurs, they together become metallic, but the lighter particles are dominant. Only at the limit $\Delta t=1$, the heavier particles are always localized.

Similarly to the region of $\Delta t>0$, when $\Delta t <0$, the mass imbalance also induces a population imbalance between two particle species, but this population imbalance is not strong as in the region of $\Delta t>0$. The band filling of the single-component particles, which become lighter with increasing $|\Delta t|$, is larger than the one of the two-component particles. At the limit of $\Delta t=-1$, the two-component particles are localized due to vanishing of their hopping amplitude. We also refer to the MIT driven by the mass imbalance in the $\Delta t<0$ region  as a species-selective-like MIT, because the lighter particles are dominant over the heavier ones in the transition. In contrast to the orbital-selective MIT, where particles of both spins always participate in the transition~\cite{Anisimov,Koga,Liebsch2004,Medici}, in this $\Delta t<0$ region, particles of a single component are dominantly involved in the MIT. It is similar to the Mott-like MIT of the spinless electrons in the Falicov-Kimball model~\cite{FreericksZlatic,Dongen92,Dongen}. The MIT occurs when the band of the single-component particles is split by a gap due to the interspecies interaction.

\begin{figure}[b]
\includegraphics[width=0.45\textwidth]{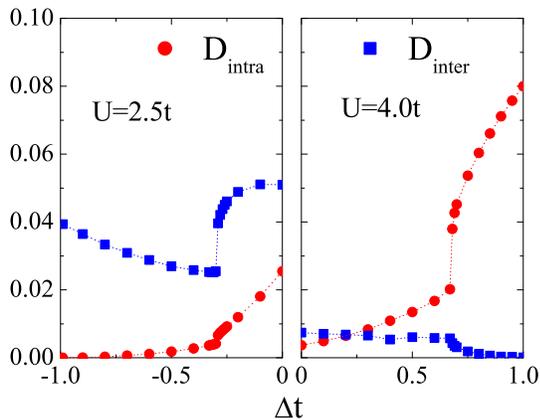}
\caption{(Color online)The intraspecies double occupancy $D_{\rm intra}$  (red filled circles)
and the interspecies double occupancy $D_{\rm inter}$ (blue filled squares) at fixed total density $n=1$ ($T=0.02 t$). The left panel plots the region of $\Delta t<0$ ($U= 2.5 t)$, while the right panel plots the region of $\Delta t>0$ ($U=4 t$).  } \label{fig3}
\end{figure}

\begin{figure}[b]
\includegraphics[width=0.34\textwidth]{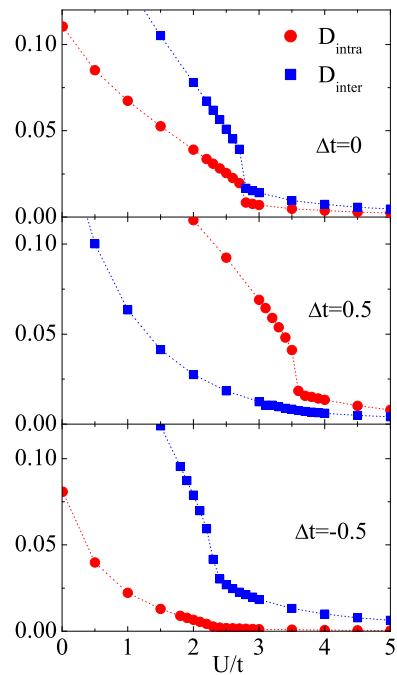}
\caption{(Color online) The intraspecies double occupancy $D_{\rm intra}$  (red filled circles)
and the interspecies double occupancy $D_{\rm inter}$ (blue filled squares) at fixed total density $n=1$ and different mass imbalances ($T=0.02 t$).}
\label{fig4}
\end{figure}

Experiments could observe the mass-imbalance-driven MIT by measuring the doubly occupied sites. Indeed,
the MIT of ultracold $^{40}$K  atoms was detected by measuring the double occupancy~\cite{Jordens}.
In Fig.~\ref{fig3} we plot the intra- and interspecies double occupancies at total particle density $n=1$ in both $\Delta t>0$ and $\Delta t<0$ regions. One can see a kink of these double occupancies at the point of the MIT. It indicates that the MIT is a first-order phase transition. In the region of $\Delta t>0$, the right panel of Fig.~\ref{fig3} shows us that the double occupancies in the insulating phase are small, but nonzero, as we have discussed previously. With increasing the mass imbalance, at the MIT, the intraspecies double occupancy abruptly increases, while the interspecies double occupancy decreases to zero. Actually, with large mass imbalances the metallic state is dominantly occupied by the two-component particles. As a consequence, the intraspecies double occupancy increases, while the interspecies double occupancy tends to vanish when $\Delta t \rightarrow 1$. In contrast to the region of $\Delta t>0$, in the region of $\Delta t<0$ (the left panel of Fig.~\ref{fig3}),
the intraspecies double occupancy decreases to zero, when $|\Delta t|$ increases. It is clear that in this region, the two-component particles become heavier, and in the $\Delta t \rightarrow -1$ limit, they are actually localized. Consequently, the intraspecies double occupancy
tends to vanish when $\Delta t \rightarrow -1$. However, the interspecies double occupancy remains finite in the insulating phase. Actually, in the region of $\Delta t <0$, the mass imbalance drives the mixture from the metallic to the insulating state. Therefore, the MIT may occur only for the intermediate local interaction that should be smaller than the critical value of the local interaction for the MIT in the mass balanced case. The local interaction in such value range is not strong enough to suppress the double occupation. Instead, it allows virtual hoppings that produce double occupation of interspecies particles, despite that the band of the single-component particles already opens a gap.

To characterize the influence of  the correlation-driven MIT in the case of mass imbalance, in Fig.~\ref{fig4} we show the dependence of the double occupancies on the local interaction for different $\Delta t$. In the balanced mass case ($\Delta t=0$), $D_{\rm inter}=2 D_{\rm intra}$. Both inter- and intra-species double occupancies abruptly change at the MIT, and they are suppressed in the insulating state. In the presence of the mass imbalance, only the intra- or interspecies double occupancy exhibits a kink at the MIT. In the region of $\Delta t>0$, the double occupancy of the two-component particles also indicates the rapid suppression at the MIT, while the interspecies double occupancy continuously decreases with increasing the local interaction. These behaviors interchange each other in the opposite region with $\Delta t<0$. They indicate the active role of the two-component particles in the $\Delta t>0$ region and of the single-component particles in the region of $\Delta t<0$ in the MIT. It means that only lighter particles are actively involved in the MIT. In the insulating phase, the double occupancies are suppressed, but they remain finite and only vanish at the strong interaction limit.
The double occupancies are experimentally accessible, but it is a challenge to experimental observations of the kink of the double occupancies.

\begin{figure}[t]
\includegraphics[width=0.38\textwidth]{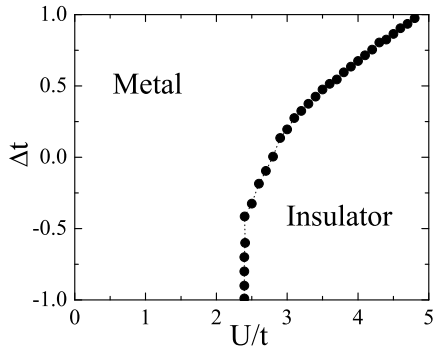}
\caption{Phase diagram at total particle density $n=1$ ($T=0.02t$).}
\label{fig5}
\end{figure}

In Fig.~\ref{fig5} we summarize a phase diagram expressing the MIT in the ($\Delta t$, $U$) plane at the commensurate total particle densities. The phase diagram is constructed from the dependence of the total particle density on the chemical potential as discussed before. The insulating phase is detected when a plateau appears in the line of $n(\mu)$.
In contrast to the two-component Hubbard model, we do not observe any phase coexistence of metal and insulator~\cite{Rozenberg,Georges}.
By increasing $\Delta t$ the critical $U$ value monotonically increases. However, since $-1 \leq \Delta t \leq 1$,
the MIT driven by the mass imbalance can occur only at a finite moderate range of the local interaction. For weak and strong local interactions, the
mass imbalance cannot drive the mixture out of its ground state.
In fermion mixtures of $^{40}$K and $^{6}$Li atoms,  $\Delta t$ can vary from $0.3$ to $0.85$~\cite{Dalmonte}.
One may expect to detect the mass-imbalance driven MIT in such ultracold atom mixtures
at commensurate particle densities and moderate local interactions.

\section{Conclusion}

We have studied the MIT driven by the mass imbalance in the three-component Hubbard model. Within the DMFT with exact diagonalization, we have found
the MIT driven by the mass imbalance at commensurate total densities, like in the balanced three-component Hubbard model~\cite{Gorelik}. The MIT is solely driven by the mass imbalance. The positive mass imbalance can only drive the mixture from insulator to metal, while the negative one drives the mixture from metal to insulator.
In order to explore the MIT in the three-component Hubbard model with mass imbalance, we have also calculated the double occupancies of both two-component particles ($D_{\rm intra}$) and different species particles ($D_{\rm inter}$) as functions of $\Delta t$ and local interaction $U$. At a critical $U$ value, the double occupancies exhibit a kink, indicating the MIT transition of the first order. The more enhancing the mass imbalance, the MIT takes place at larger critical $U$ value. The phase diagram expressing the MIT in the ($\Delta t$, $U$) plane is also constructed. It shows that the MIT occurs only
at moderate local interactions. For weak and strong local interactions, the mass imbalance cannot drive the system out of its ground state.
Actually, the mass imbalance induces a population imbalance between particle species, and the lighter particles dominantly take part in the MIT.
The MIT can be interpreted as
a light-particle species selective transition. These features are distinct from the behaviors of balanced systems.
We predict the MIT driven by the mass imbalance in fermion-fermion mixtures of ultracold atoms, loading into optical latices,
for instance, the mixture of $^{40}$K and $^{6}$Li atoms. By measuring the ratio of the doubly occupied sites in the optical lattices,
the MIT could be observed by tuning the mass imbalance ratio.

\section*{Acknowledgement}

We thank Le Duc-Anh for useful discussions.
This research is funded by Vietnam National Foundation
for Science and Technology Development (NAFOSTED) under Grant No. 103.01-2014.09.

\end{document}